\documentclass[amsmath,amssymb,reprint]{revtex4-1}
\usepackage{graphicx}
\usepackage{graphics}
\usepackage{color}
\usepackage{ulem}
\usepackage{version} 

\usepackage{bm}
\usepackage{textcomp}

\DeclareMathAlphabet{\mathit}{OT1}{ptm}{m}{it}

\def \be {\begin{equation}}
\def \ee {\end{equation}}
\def \bea {\begin{eqnarray}}
\def \eea {\end{eqnarray}}

\def \bd  {\begin{details}}
\def \ed  {\end{details}}

\begin{document}

\excludeversion{details}


\title{Texture-induced spin-orbit coupling and  
Skyrmion-electron bound states \\ 
in a N\'eel antiferromagnet}

\author{N. Davier}
\email[]{davier@irsamc.ups-tlse.fr}
\affiliation{Laboratoire de Physique Th\'eorique, Universit\'e de Toulouse, CNRS, UPS, France}

\author{R. Ramazashvili}
\email[]{revaz@irsamc.ups-tlse.fr}
\affiliation{Laboratoire de Physique Th\'eorique, Universit\'e de Toulouse, CNRS, UPS, France}

\date{\today}

\begin{abstract} 
We derive effective-mass electron Hamiltonian for a N\'eel antiferromagnet
in the presence of a smooth texture of the staggered magnetization. 
For certain locations of electron band extrema, the texture 
produces a peculiar and anomalously strong spin-orbit coupling 
of the scale $\hbar v / L$, with $v$ the Fermi velocity and $L$ 
the characteristic length scale of the texture. 
For a Skyrmion texture, this coupling generates electron 
bound states, whose energy scale is given by the gap $\Delta$ in the 
electron spectrum. With dopant carriers, such bound states turn the 
Skyrmion into a charged particle, that can be manipulated by electric field. 
\end{abstract}

\maketitle

\section{Introduction} 
\label{section:intro}

It is an honor and our great pleasure to contribute to this Festschrift 
for the 95-th birthday of Emmanuel Rashba. One of us (R.R.) first came 
under the spell of E. I. Rashba's papers as a student, and turns to them 
for understanding and inspiration to this day. This modest offering 
is our `thank you'. Happy birthday, Emmanuel Iosifovich, 
many happy returns of the day!

Topological textures such as domain 
walls, vortices and Skyrmions appear prominently in diverse areas 
of physics, from cosmology and string theory \cite{Schwarz} to QCD, 
the physics of hadrons, and condensed matter physics \cite{Rho}. 
In solid state magnetism alone, topological textures bring together 
fundamental and applied science, from novel states of matter such as 
Skyrmion crystal to prototype spintronic devices that employ Skyrmions 
and domain walls to process information \cite{Back-2020}. 
While early spintronics research largely focussed on ferromagnetic 
materials \cite{Fert-2008,Gruenberg-2008,Fert-2019}, 
an ever increasing effort has been turning to antiferromagnets \cite{Gomonay-2017,Baltz-2018} and to topological textures 
therein \cite{Smejkal-2018,Goebel-2021} -- in view of their 
technologically attractive properties such as shorter characteristic 
timescales of the N\'eel state and its lack of net magnetization 
\cite{Jungwirth-2016}. 

In order to usefully employ topological textures, it is crucial 
to understand their interplay with other subsystems of a solid, 
first and foremost -- with band electrons. For ferromagnets, 
much work has been done to understand the influence of a texture 
on electric current and vice versa \cite{Tatara-2008,Nagaosa-2013}.
Soon after, this effort has been expanded to antiferromagnets 
\cite{MacDonald-2011,Gomonay-2014,Baltz-2018,Manchon-2019}. 

An important step in this direction has been undertaken by R. Cheng and 
Q. Niu \cite{Cheng-2012}, who studied electron motion in the presence of 
a texture in a N\'eel antiferromagnet, using coupled quasiclassical 
equations of motion for the electron momentum, coordinate and spin. 
Here we address the same problem fully quantum-mechanically, 
by deriving the effective-mass Hamiltonian 
 -- and show how, for certain locations of electron band extrema, 
the texture generates a peculiar and anomalously strong spin-orbit 
coupling, a key result of this work.  

It is at this point that we find ourselves following in the footsteps 
of Solomon Pekar and Emmanuel Rashba \cite{Pekar-1964}. 
Back in 1964, Pekar and Rashba observed that a non-uniform magnetic 
field as well as non-uniform magnetization $\mathcal{M} ({\bf r})$ 
give rise to a spin-orbit coupling: exchange interaction 
$\mathcal{M} ({\bf r}) \cdot {\bm \sigma}$ 
couples the electron coordinate ${\bf r}$ to its spin ${\bm \sigma}$ 
via ${\bf r}$-dependence of $\mathcal{\bm M} ({\bf r})$ -- that is, 
simply by virtue of inhomogeneity. 

The observation above appears to suggest non-degenerate bands, 
split by exchange $\mathcal{M} ({\bf r}) \cdot {\bm \sigma}$. Yet 
symmetry can void this argument by restoring the double degeneracy, 
as it does for a centrosymmetric N\'eel antiferromagnet in its uniform state: 
here, magnetization $\mathcal{\bm M} ({\bf r})$ changes sign 
upon translation ${\bf T}_a$ by a lattice period $a$ as well as 
upon time reversal $\theta$. As a result, combined anti-unitary 
symmetry $I \theta {\bf T}_a$, with $I$ the inversion operator, 
guarantees double degeneracy of Bloch eigenstates throughout 
the Brillouin zone \cite{Herring}.

A localized texture breaks such a degeneracy-protecting symmetry. 
Below we illustrate this by a texture in a N\'eel antiferromagnet, 
and by an unusual spin-orbit coupling that it produces. 
Moreover, for a Skyrmion texture, this spin-orbit coupling creates 
electron bound states with energy scale given by the gap $\Delta$ 
in the electron spectrum. With dopant carriers, such bound states 
turn the Skyrmion into a charged particle, another key result of this work.
 
The paper is organized as follows. In Section \ref{section:texture}, 
we begin by deriving the low-energy electron Hamiltonian for a 
centrosymmetric N\'eel antiferromagnet in the presence of a 
texture. Then we specify the Hamiltonian for a particular location 
of the electron band extrema -- and point out the appearance of 
a peculiar spin-orbit coupling. 
In Section \ref{section:BPSkyrmion}, we present our main 
example: a texture in the form of a single Belavin-Polyakov 
Skyrmion \cite{Belavin-1975,Rajaraman}. We show that the 
texture-induced spin-orbit coupling produces Skyrmion-electron bound 
states, and study their evolution as a function of the Skyrmion radius. 
In Section \ref{section:discussion}, we discuss the implications 
of our results in the light of the pioneering work 
\cite{Pekar-1964} of Pekar and Rashba. 
Finally, the Appendices discuss the connection of our results 
with some of the earlier studies, and outline the validity range 
of the approximations we used.  

\section{Electron in the presence \\ of a texture} 
\label{section:texture}

Consider a N\'eel antiferromagnet on a square-symmetry lattice with 
\mbox{period $a$}. In the uniform state, its ordered moment changes 
sign upon elementary translation, and couples electron states at any two 
momenta ${\bf p}$ and ${{\bf p} + {\bf Q}}$, separated by the N\'eel 
wave vector ${\bf Q} = (\pm \frac{\pi}{a} , \pm \frac{\pi}{a})$. 
The coupling has the form of exchange $({\bm \Delta} \cdot {\bm \sigma})$, 
with ${\bm \Delta}$ proportional to the staggered magnetization, and 
${\bm \sigma}$ the triad of Pauli matrices, representing electron spin. 
Since ${\bf p}$ and ${\bf p} + 2 {\bf Q}$ are equivalent in the Brillouin zone (BZ), 
the Hamiltonian $\mathcal{H}$ can be written as acting on a bispinor 
$\Psi = (\psi_{\bf p} , \psi_{\bf p+Q})$ \cite{KulTug.1984}:
\be
\label{eq:H4x4-uniform}
\mathcal{H} = 
\left[
\begin{array}{cc}
\varepsilon ({\bf p})  &
({\bm \Delta} \cdot {\bm \sigma}) \\
( {\bf \Delta} \cdot {\bm \sigma}) &
\varepsilon ({\bf p + Q}) 
\end{array}
\right] ,
\ee 
where $\varepsilon ({\bf p})$  is the electron dispersion 
in the absence of N\'eel order. The spectrum 
$E_{\bf p}$ of $\mathcal{H}$ is doubly-degenerate, 
$E_{\bf p} = \varepsilon_{+}({\bf p}) \pm \sqrt{|{\bm \Delta}|^2 + \varepsilon_{-}^2 ({\bf p})}$,   
where $\varepsilon_{\pm}({\bf p}) \equiv \frac{1}{2} 
\left[ \varepsilon ({\bf p}) \pm \varepsilon ({\bf p + Q}) \right]$;  
it has a gap $\Delta = | {\bm \Delta} |$, 
which turns a half-filled metal into an insulator. 

In the presence of a texture ${\bm \Delta}_{\bf r} = \hat{\bf n}_{\bf r} \Delta$, 
unit vector $\hat{\bf n}_{\bf r}$ becomes a smooth function of the coordinate 
${\bf r}$, and carriers near the extrema of $E_{\bf p}$ at momenta ${\bf p}_0$ 
and ${{\bf p}_0  + {\bf Q}}$ admit a low-energy effective-mass description \cite{Kittel}.  
To derive it, in Eq. (\ref{eq:H4x4-uniform}) we replace uniform ${\bm \Delta}$ 
by ${\bm \Delta}_{\bf r}$, and substitute $\hat{\bf p} \equiv - i \hbar {\bm \nabla}$ 
for the momentum dependences $\varepsilon_{{\bf p}_0} (\hat{\bf p})$ 
and $\varepsilon_{{\bf p}_0 + {\bf Q}} (\hat{\bf p})$ of 
$\varepsilon ({\bf p})$ near ${\bf p}_0$ and ${\bf p}_0  + {\bf Q}$. 

Now perform a spin rotation $U_{\bf r}$ that makes ${\bm \Delta}_{\bf r}$ 
uniform: $U_{\bf r}^\dagger (\hat {\bf n}_{\bf r} \cdot {\bm \sigma}) U_{\bf r} = 
 \sigma_z$ \cite{Volovik-1987}. This generates a Peierls 
 substitution $\hat{p}_i \rightarrow \hat{p}_i + ({\bf A}_i \cdot {\bm \sigma})$ 
 in $\varepsilon_{{\bf p}_0} (\hat{\bf p})$ and 
$\varepsilon_{{\bf p}_0 + {\bf Q}} (\hat{\bf p})$, 
with $({\bf A}_i \cdot {\bm \sigma})
 = A_i^\alpha \sigma_\alpha = - i \hbar U_{\bf r}^\dagger \partial_i U_{\bf r}$. 
Vector potential $A_i^\alpha$ carries real-space indices $i = x , y$ 
and spin indices $\alpha = x , y , z$. 
While different components of $({\bf A} \cdot {\bm \sigma})$ 
do not commute, $U_{\bf r}$ is defined only up to a non-uniform 
spin rotation $V^z_{\bf r} = e^{i \sigma_z \chi}$ around 
$\hat{z}$: $U_{\bf r} \rightarrow U_{\bf r} V^z_{\bf r}$, 
which is an abelian transformation. 
This gauge transformation acts  on $({\bf A} \cdot {\bm \sigma})$ 
in a peculiar way, elucidated by first-order expansion in infinitesimal $\chi$:  
\be
\label{eq:gauge-transformation}
\delta ({\bf A}_i \cdot {\bm \sigma}) 
= \hbar \sigma_z \partial_i \chi
+ \chi 
\left[ 
({\bf A}_i \cdot {\bm \sigma}) , \sigma_z
\right] .
\ee
That is, $A_i^z$ transforms as electromagnetic vector potential 
($\delta A_i^z = \hbar \partial_i \chi$), while ${\bf A}_i^\| = (A_i^x , A_i^y)$ 
rotates around $\hat{z}$ by angle $2 \chi$. 
This observation will prove useful below.

Next, we split the bispinor $\Psi$ 
into two spin-$\frac{1}{2}$ components, for states at energies near 
$\pm \Delta$, respectively -- and thus take the $4 \times 4$ (`Dirac') 
Hamiltonian (\ref{eq:H4x4-uniform}) to its $2 \times 2$ (`Pauli-Schr\"odinger') 
low-energy limit \cite{LL-IV,Ryder}. 
Here, we focus on the conduction band (energies $E$ near $+\Delta$) 
and, to first order in ${\frac{E - \Delta}{\Delta} \ll 1}$, find the effective-mass 
Hamiltonian $\mathcal{H}_{{\bf p}_0}$ near ${\bf p}_0$,  
with $\bar\varepsilon_{{\bf p}_0 + {\bf Q}} (\hat{\bf p}) \equiv 
\sigma_z \varepsilon_{{\bf p}_0 + {\bf Q}} (\hat{\bf p}) \sigma_z$:  
\be
\label{eq:low-E-Hamiltonian-general}
\mathcal{H}_{{\bf p}_0} = 
\frac{
\varepsilon_{{\bf p}_0} (\hat{\bf p})  
+ \bar\varepsilon_{{\bf p}_0 + {\bf Q}} (\hat{\bf p}) 
}{2}
 + 
\frac{ 
 \left[
\varepsilon_{{\bf p}_0} (\hat{\bf p}) 
  - 
\bar\varepsilon_{{\bf p}_0 + {\bf Q}} (\hat{\bf p}) 
 \right]^2
 }{8 \Delta} .
\ee
\begin{figure}[tb] 
    \centering
        \includegraphics[width=0.25\textwidth]{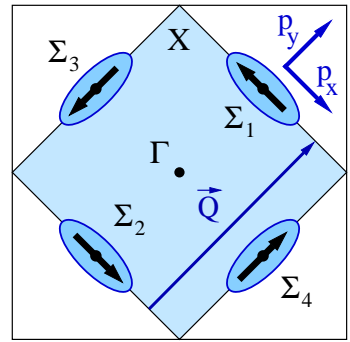}
            \caption{
The Brillouin zone (BZ) of a square-lattice N\'eel antiferromagnet 
with wave vector $\vec{\bf Q} = (\pm \frac{\pi}{a} , \pm \frac{\pi}{a})$. 
The large square shows the BZ in the paramagnetic state, the shaded 
square depicts the Brillouin zone in the N\'eel state (MBZ). The band 
extrema are assumed to lie at face centers $\Sigma_1$-$\Sigma_4$ 
of the MBZ. The $p_x , p_y$ are the local momentum axes near $\Sigma_1$, 
as used in the main text. The ellipses sketch the equal-$E_{\bf p}$ 
lines near $\Sigma_1$-$\Sigma_4$. The bold arrows centered at 
$\Sigma_1$-$\Sigma_4$ show the electron spin polarization 
of low-energy conduction-band bound states in each valley, 
for a large BP Skyrmion of N\'eel type (see main text).
}
\label{Fig:BZ}
\end{figure}
The explicit form of Hamiltonian (\ref{eq:low-E-Hamiltonian-general}) 
depends on that of   $\varepsilon_{{\bf p}_0} (\hat{\bf p})$  and 
$\varepsilon_{{\bf p}_0 + {\bf Q}} (\hat{\bf p})$,  in its turn defined 
by the sym\-metry of momenta  ${\bf p}_0$  and  ${\bf p}_0 + {\bf Q}$ 
in the BZ. Hereafter we focus on the extrema at midpoints $\Sigma$ 
of the magnetic Brillouin zone (MBZ) boundary in Fig. \ref{Fig:BZ}. 
At point $\Sigma$, the momentum expansion of 
$\varepsilon_{{\bf p}_0} (\hat{\bf p})$ and 
$\varepsilon_{{\bf p}_0 + {\bf Q}} (\hat{\bf p})$ begins with 
$\pm {\bf v} \cdot \hat{\bf p} + \hat{p}_i^2 / 2m_i$, with the 
paramagnetic-state Fermi velocity ${\bf v}$ at $\Sigma$  
pointing along the local $p_y$ in Fig. \ref{Fig:BZ}, 
and $m_i = m_x , m_y$ the paramagnetic-state 
effective masses along and normal to the MBZ boundary.  
Truncating the momentum expansion of Eq. 
(\ref{eq:low-E-Hamiltonian-general}) at quadratic terms 
\cite{footnote:re-Hamiltonian}, we find 
\be 
\label{eq:low-E-Hamiltonian-point-X}
\mathcal{H}_\Sigma
 = \frac{\left(\hat{p}_i + A^z_i \sigma_z \right)^2}{2m_i^*} 
  + \frac{\left( A^\|_i \right)^2}{2m_i} 
  + v \left( {\bf A}_y^\| \cdot  {\bm \sigma} \right) .
\ee

Hamiltonian (\ref{eq:low-E-Hamiltonian-point-X}) 
\textit{is} gauge-invariant: as per Eq. (\ref{eq:gauge-transformation}), the 
$A^z_i$ in the kinetic energy transforms as electromagnetic vector potential, 
while ${\bf A}_i^\| = (A_i^x , A_i^y)$ in the remaining terms transforms 
by rotation around $\hat{z}$. Moreover, the terms above are the only ones 
allowed by symmetry to second order in momentum, keeping in mind  
symmetry under reflection $x \rightarrow -x$ \cite{footnote:x-reflection}.  
The astute reader will also notice that the second term in Hamiltonian 
(\ref{eq:low-E-Hamiltonian-point-X}) is small relative to the third one as long 
as the characteristic length scale of the texture remains large compared with 
the lattice spacing, that is as long as the continuum description of the texture 
applies. The concrete problem we treat in Section \ref{section:BPSkyrmion} 
confirms this observation.

The non-uniformity of a texture appears in Hamiltonian 
(\ref{eq:low-E-Hamiltonian-point-X}) via the vector potential 
$({\bf A} \cdot {\bm \sigma})$ that couples the electron spin to 
its orbital motion. However, obeying different gauge transformation rules as 
per Eq. (\ref{eq:gauge-transformation}), the $z$-component $A^z_i \sigma_z$ 
and the `in-plane' spin components $({\bf A}_y^\| \cdot  {\bm \sigma})$ 
produce spin-orbit coupling differently:  
The former induces `gauge' spin-orbit coupling via the 
Peierls substitution in the kinetic energy, while 
$v \left( {\bf A}_y^\| \cdot  {\bm \sigma} \right)$ 
acts as a texture-induced Zeeman field. 
In Section \ref{section:discussion}, we will discuss the 
peculiarities of this spin-orbit coupling in relation 
to the work \cite{Pekar-1964} of Pekar and Rashba. 

Finally, note that $m_y^*$ in Eq. (\ref{eq:low-E-Hamiltonian-point-X})  
is renormalized relative to $m_y$ in the expansion of 
$\varepsilon_{{\bf p}_0} (\hat{\bf p})$ and 
$\varepsilon_{{\bf p}_0 + {\bf Q}} (\hat{\bf p})$, as per 
$(m_y^*)^{-1} = ( m_y)^{-1} + \frac{v^2}{\Delta}$. 
Therefore,  $m_y^*$ is small against $m_y$: 
$\frac{m_y^*}{m_y} \sim \frac{\Delta}{\epsilon_F} \ll 1$, 
while $m_x^* = m_x$ is of the order of band electron mass $m$ 
or greater \cite{footnote:m_x}. 
Such a mass anisotropy arises for \textit{any} (not only N\'eel) 
$(\frac{\pi}{a} , \frac{\pi}{a} )$ order with a gap $\Delta \ll \epsilon_F \equiv m_y v^2$. 
Electron-doped cuprates at low-to-optimal doping provide a prominent example  
\cite{Armitage,Matsui,Song}, even if the nature of their ordering remains 
controversial, with experimental evidence presented both against 
\cite{luke90,moto07,mang04a,saad15} and for 
\cite{yama03,kang05,daga05, yu07,dora18,RR-2021} 
the presence of (quasi)static N\'eel order. 

\section{A tractable example: Belavin-Polyakov Skyrmion}
\label{section:BPSkyrmion} 

Having obtained Hamiltonian (\ref{eq:low-E-Hamiltonian-point-X}),
let us turn to a concrete problem: the $({\bf A} \cdot {\bm \sigma})$ 
defined by a single Skyrmion. 
Consider a centrosymmetric isotropic antiferromagnet with stiffness $J$ 
and continuum-limit energy density $J ({\bm \nabla} \hat{\bf n}_{\bf r})^2$. 
In the topological sector with winding number $\mathcal{Q} = 0, \pm 1, \pm 2 ...$, 
the lowest-energy solution is the Belavin-Polyakov (BP) Skyrmion 
\cite{Belavin-1975,Rajaraman} defined by a single length scale: the radius $R$. 
The energy $4 \pi J | \mathcal{Q} |$ of the BP Skyrmion is 
independent of $R$ by virtue of scale invariance of energy 
$J \int d^2{\bf r} ({\bm \nabla} \hat{\bf n}_{\bf r})^2$. 
Being defined by a single length scale makes the BP Skyrmion the 
simplest case to analyze, which leads us 
to study Hamiltonian (\ref{eq:low-E-Hamiltonian-point-X}) for a $\mathcal{Q} = 1$ 
BP Skyrmion. 
We focus on a 
configuration $\hat{\bf n}_{\bf r} = (\sin \theta \cos \phi , \sin \theta \sin \phi , \cos \theta)$ 
with the polar angle $\theta$ 
depending only on the distance $r = \sqrt{x^2 + y^2}$ to the Skyrmion center, 
and $\phi = \arctan \frac{y}{x}$ being the azimuthal angle in the $(x,y)$ plane. 
The energy density $J ({\bm \nabla} \hat{\bf n}_{\bf r})^2$ of such a configuration 
is invariant under shifting $\phi$ by a constant $\gamma$, usually called `helicity' \cite{Goebel-2021}. Moreover, for the BP Skyrmion, the helicity drops out 
of the electron problem up to the direction of the resulting spin polarization. 
This allows us to reduce the setting to the $\gamma = 0$ 
pattern above, commonly called the `N\'eel' Skyrmion. 

Note that the latter has its own localized eigenexcitations \cite{Kravchuk-2019}, 
which, generally, shall be treated on an equal footing with the electron degrees 
of freedom. However, for a sufficiently small single-ion anisotropy and a not too 
large BP Skyrmion, the proper Skyrmion frequencies are small against those 
of electron motion (see Appendix \ref{section:static}). 
In this limit, treating the electron problem as if the Skyrmion were 
perfectly static is a reasonable first approximation that we now focus on. 

To proceed, we need to fix the gauge, that is to select a concrete $U_{\bf r}$. 
We do so by choosing 
\be
\label{eq:U}
U_{\bf r} = ({\bf m}_{\bf r} \cdot {\bm \sigma}) ,
\ee 
with unit vector ${\bf m}_{\bf r}$ pointing along the bisector between 
$\hat{z}$ and ${\bm \Delta}_{\bf r}$. Being equivalent to $\pi$-rotation 
around ${\bf m}_{\bf r}$, such a $U_{\bf r}$ brings ${\bm \Delta}_{\bf r}$ 
to point along $\hat{z}$ \cite{Tatara-2008}. 

The $\mathcal{Q} = 1$ BP Skyrmion profile is 
$\sin \theta = \frac{2z}{1 + z^2}$ with $z = \frac{r}{R}$ 
\cite{Belavin-1975,Rajaraman} (in passing, notice that 
$\theta \left[ R\right] = \frac{\pi}{2}$). 
In the chosen gauge, calculation of spin-$z$ components $A^z_i$ 
in the first term of Eq. (\ref{eq:low-E-Hamiltonian-point-X}) yields 
\be
\label{eq:Az}
A^z_x = 
\frac{- \hbar y}{R^2 + r^2} \,\, , \,\,
A^z_y = 
\frac{\hbar x}{R^2 + r^2} .
\ee
Thus, the Skyrmion produces geometric flux $\pm 2 \pi \hbar$ 
for the spin-up and spin-down components of the wave function: 
$\oint A^z_i dl_i = 2 \pi \hbar$, with the integral taken along a large 
contour of radius $r \gg R$ \cite{footnote:flux}. Such a flux induces 
topological spin Hall effect \cite{Yin-2015,Buhl-2017,Akosa-2018}. 

The last term in Hamiltonian (\ref{eq:low-E-Hamiltonian-point-X}) takes the form  
\be
\label{eq:vF}
v \left( {\bf A}_y^\| \cdot  {\bm \sigma} \right) = 
- \frac{\hbar v}{R} \frac{\sigma_x}{1 + z^2} 
 = 
 - \Delta \frac{\xi}{R} \frac{\sigma_x}{1 + z^2} 
 , 
\ee 
where, by analogy with superconductivity,  we choose to call 
$\xi = \hbar  v / \Delta$ the antiferromagnetic coherence length. 
This term couples the electron spin to its orbital motion and 
produces an attractive potential for the spin-up component of 
the wave function along the $\hat{x}$ axis \cite{footnote:BP}.  

Finally, the second term in Eq. (\ref{eq:low-E-Hamiltonian-point-X}) 
creates a repulsive potential 
\be
\label{eq:centrifugal} 
\frac{\left( A^\|_i \right)^2}{2m_i}  = 
\frac{\hbar^2}{2 R^2} 
\left[ \frac{1}{m_x} + \frac{1}{m_y} \right] 
\frac{1}{(1 + z^2)^2}.  
\ee 
Comparing the terms (\ref{eq:centrifugal}) and (\ref{eq:vF}), 
we see that the latter is an $a/R \ll 1$ fraction of the former. 
That is, the repulsion (\ref{eq:centrifugal}) is negligible relative to 
the spin-orbit term (\ref{eq:vF}) as long as the continuum description 
of the Skyrmion is valid.

The term (\ref{eq:vF}) is precisely the `texture-induced Zeeman' 
part of the spin-orbit coupling that we discussed below 
Eq. (\ref{eq:low-E-Hamiltonian-point-X}). 
Direct inspection shows that, for a large (${R \gg \xi}$) Skyrmion, the r.h.s. 
of Eq. (\ref{eq:vF}) overwhelms all the other terms with $A^\alpha_i$ 
in Eq. (\ref{eq:low-E-Hamiltonian-point-X}) \cite{footnote:small}, 
and creates non-degenerate Skyrmion-electron bound states, 
a key result of our work. At $R \gg \xi$, the low-lying bound 
states are shallow and spin-polarized in each of the four $\Sigma$ 
valleys as shown in Fig. \ref{Fig:BZ} \cite{footnote:chirality}. 

Note that this polarization is opposite to the one expected from Eq. 
(\ref{eq:vF}). This is a result of undoing the transformation $U_{\bf r}$ 
of Eq. (\ref{eq:U}) to restore the original spin axes. While $U_{\bf r}$ 
is substantially non-uniform, the reader will see that, for $R \gg \xi$, 
it does remain nearly constant over the spatial extent of the low-lying 
 bound states: $U_{\bf r} = ({\bf m}_{\bf r} \cdot {\bm \sigma}) \approx \sigma_z$. 
That is, for low-lying bound states, undoing the $U_{\bf r}$ amounts 
to a spin rotation by $\pi$ around $\hat{z}$, which simply inverts the 
spin polarization relative to the one dictated by Eq. (\ref{eq:vF}). 
Notice that for smaller Skyrmions low-lying bound states are no 
longer uniformly spin-polarized: instead, their spinor structure 
varies substantially over the wave function range. 

 \begin{figure}[tb] 
    \centering
        \includegraphics[width=0.35\textwidth]{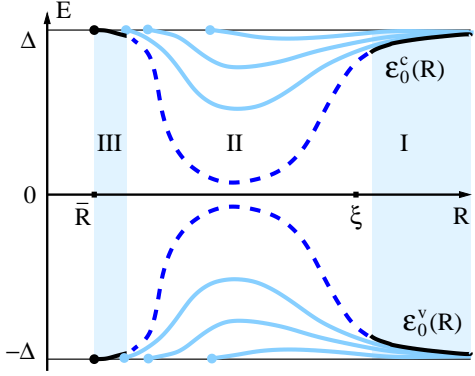}
            \caption{
Energy $\epsilon_0^c (R)$ of BP Skyrmion-electron bound state, 
ge\-ne\-ra\-ted by the conduction band, and of its valence-band 
counterpart $\epsilon_0^v (R)$ (the highest filled state), 
sketched as a function of the BP Skyrmion radius $R$. 
\textbf{In region I} ($R \gg \xi = \frac{\hbar v}{\Delta}$), 
the bound state is described by low-energy Hamiltonian 
(\ref{eq:low-E-Hamiltonian-point-X}). 
The $\epsilon_0^c (R)$ is given by Eq. (\ref{eq:low-E-at-large-R}) 
and shown by solid line. 
\textbf{In region II} ($\bar{R} \ll R \lesssim \xi$), 
the low-energy approximation breaks down, and the bound state 
(dashed line) must be found from the full Hamiltonian of a kind 
(\ref{eq:H4x4-uniform}) in the presence of the Skyrmion, 
which goes beyond the scope of this work. 
\textbf{In region III} (narrow range ${R - \bar{R} \ll \bar{R}}$), the bound 
state, shown by solid line, becomes shallow again.
Its disappearance at  $R = \bar{R}$ can be described by low-energy 
Hamiltonian (\ref{eq:low-E-Hamiltonian-point-X}) (see main text). 
Pale lines sketch the higher bound states. 
}
\label{Fig:Scales}
\end{figure}

The mass anisotropy $\frac{m_y^*}{m_x^*} \sim \frac{\Delta}{\epsilon_F} \ll 1$ 
of Hamiltonian (\ref{eq:low-E-Hamiltonian-point-X}) makes the $y$ coordinate 
`fast' re\-la\-tive to $x$, and the energies of low-lying bound states can be 
readily evaluated in the Born-Oppenheimer approximation \cite{Tully}. 
For $R \gg \xi$, the energies $\epsilon_n^c$ of the bound states generated 
by the conduction band and labelled by quantum number $n$ can be 
evaluated by expanding the r.h.s. of Eq. (\ref{eq:vF}) to first order in $z^2$ and 
finding the spectrum of the ensuing harmonic oscillator with respect to $y$: 
\be
\label{eq:low-E-at-large-R}
\epsilon_n^c (R) 
\approx 
  - \Delta\frac{\xi}{R} 
\left[ 
1 - \sqrt{\frac{2\xi}{R}} 
\left( n + \frac{1}{2} \right)
\right] .
\ee
Subsequent account of the `slow' coordinate $x$ generates 
energy levels that we label with quantum number $k$, with 
the effective oscillator frequency of the order of 
${\sqrt{\frac{m_y}{m_x} \frac{a}{\xi}} \ll 1}$ relative to the one above: 
\begin{widetext} 
\be
\label{eq:low-E-at-large-R-full}
\epsilon_{n k}^c (R) \approx 
  - \Delta\frac{\xi}{R} 
\left[ 
1 - \sqrt{\frac{2\xi}{R}} 
\left\{ 
\left( n + \frac{1}{2} \right)
 + 
 \sqrt{\frac{m_y}{m_x} \frac{a}{\xi}}
\left( k + \frac{1}{2} \right)
\right\}
\right] .
\ee 
\end{widetext}
Needless to say, this `fine' structure is much more prone to effects 
of proper eigenexcitations \cite{Kravchuk-2019} of the Skyrmion 
than the r.h.s. of Eq. (\ref{eq:low-E-at-large-R}), 
see Appendix \ref{section:static}. 

For $R \gg \xi$, the bound states  
(\ref{eq:low-E-at-large-R}-\ref{eq:low-E-at-large-R-full}) are shallow 
(${| \epsilon_0^c | \ll  \Delta}$), and the low-energy approximation of Hamiltonians 
(\ref{eq:low-E-Hamiltonian-general}) and (\ref{eq:low-E-Hamiltonian-point-X}) 
remains valid. However, with $R$ decreasing, $| \epsilon_0^c |$ grows 
to attain the order of $\Delta$ at $R \sim \xi$, where the low-energy 
approximation breaks down along with Hamiltonians 
(\ref{eq:low-E-Hamiltonian-general}) and (\ref{eq:low-E-Hamiltonian-point-X}), 
as sketched in Fig. \ref{Fig:Scales}. 
  
Now we will show that, with $R$ decreasing further below $\xi$, 
the bound state becomes shallow again, and vanishes at an 
$\bar{R} \sim \sqrt{\frac{m_y}{m_x}} \sqrt{\xi a}$. 
To make this length scale manifest, we eliminate $A^z_y$ 
from the first term in Eq. (\ref{eq:low-E-Hamiltonian-point-X}) 
by gauge transformation 
\be 
\label{eq:chi} 
W = e^{i \sigma_z \chi} , \,\, 
\chi \left( \tilde{x} , \tilde{y} \right) =  \frac{-\tilde{x}}{\sqrt{1 + \tilde{x}^2}} 
\arctan{ \frac{\tilde{y}}{\sqrt{1 + \tilde{x}^2}}} ,
\ee 
where $\tilde{x} = \frac{x}{R}$ and $\tilde{y} = \frac{y}{R}$. 
As a result, Hamiltonian (\ref{eq:low-E-Hamiltonian-point-X}) 
takes the form 
\be 
\label{eq:low-E-Hamiltonian-point-X-gauged}
\tilde{\mathcal{H}}_\Sigma
 =
  \frac{\hat{p}_y^2}{2m_y^*}
   + v \left( \tilde{\bf A}_y^\| \cdot  {\bm \sigma} \right)  
  + \frac{\left( A^\|_i \right)^2}{2m_i} 
  +  \frac{\left(\hat{p}_x + \tilde{A}^z_x \sigma_z \right)^2}{2m_x} ,
\ee
where $\tilde{A}^z_x = A^z_x + \partial_x \chi$, and 
$$
( \tilde{\bf A}_y^\| \cdot {\bm \sigma} ) = 
A_y^\| \left[ \sigma_x \cos 2 \chi +  \sigma_y \sin 2 \chi \right] . 
$$
In terms of the `fast' coordinate $y$, Hamiltonian 
(\ref{eq:low-E-Hamiltonian-point-X-gauged}) describes a particle 
in a one-dimensional potential $\mathcal{U}_x (y )$, parametrically 
dependent on the `slow' variable $x$, which again invites the 
Born-Oppenheimer approximation. Comparing the characteristic 
value $\frac{\hbar v}{R} = \Delta \frac{\xi}{R}$ of the second term 
in Eq. (\ref{eq:low-E-Hamiltonian-point-X-gauged}) 
with typical kinetic energy 
${\frac{\hbar^2}{m_y^* R^2} \approx \Delta \left( \frac{\xi}{R} \right)^2}$ 
of the trapped electron, we see that the former is indeed small 
against the latter for $R \ll \xi$, the range in question. 
The effective bound-state energy is thus defined by the 
integrated potential ${u (x) = \int \mathcal{U}_x ( y ) d y}$ \cite{LL-III}. 
Contribution of the second term in Eq. 
(\ref{eq:low-E-Hamiltonian-point-X-gauged}) to $u (x)$ is 
\be
u_1 (x) \sigma_x 
 = 
 - v \int d y \left( \tilde{\bf A}_y^\| \cdot  {\bm \sigma} \right)  
 =
\hbar v \sigma_x  
  \frac{
 \sin 
 \left( 
 \frac{\pi \tilde{x}}{\sqrt{1 + \tilde{x}^2}}
  \right) 
  }{\tilde{x}} .
\ee 
The third term in Eq. (\ref{eq:low-E-Hamiltonian-point-X-gauged}) 
is an $\frac{a}{R} \ll 1$ fraction of the second, thus its contribution 
to $u(x)$ can be neglected as long as continuum description 
of the Skyrmion applies ($R \gg a$). By contrast, 
\be 
u_2 (x)
 = 
\frac{1}{2 m_x} \int d y (\tilde{A}^z_x )^2 ,
\ee
arising from the last term, requires care since $(\tilde{A}^z_x )^2$ 
remains finite as $y \rightarrow \infty$: 
$$
A^2 (x) 
 \equiv 
\lim_{y \rightarrow \infty} ( \tilde{A}^z_x )^2 
 = 
\frac{\hbar^2}{R^2} 
 \frac{[ \pi/2 ]^2}{(1 + \tilde{x}^2)^3} ,
$$
which makes $u_2 (x)$ diverge. We remedy this by writing  
$( \tilde{A}^z_x )^2 
  = \left[ ( \tilde{A}^z_x )^2 - A^2 (x) \right] + A^2 (x)$. The term 
  in the square brackets then gives a finite contribution to $u (x)$, 
  which is suppressed relative to $u_1 (x)$ by a factor 
  $\frac{m_y}{m_x} \frac{a}{R}$, and hence remains negligible within  
continuum description \mbox{($R \gg a$).} The resulting bound-state 
energy $w_- (x)$ at a given $x$ is thus defined \cite{LL-III} 
by $u_1 (x)$ alone: 
\be
\label{eq:w-}
w_- (x) 
 =
  - \frac{m_y^*}{2 \hbar^2} \left[ u_1 (x) \right]^2
   = 
  - \frac{\Delta}{2} 
  \frac{
 \sin^2 
 \left(
 \frac{\pi \tilde{x}}{\sqrt{1 + \tilde{x}^2}}
 \right)
  }{\tilde{x}^2} .
\ee
It competes with the repulsive contribution of $A^2 (x)$:
 \be
 \label{eq:w+}
 w_+ (x) 
  = \frac{A^2 ( x )}{2 m_x} 
 = 
  \frac{\hbar^2}{2 m_x R^2} 
 \frac{[ \pi/2 ]^2}{(1 + \tilde{x}^2)^3} .
 \ee 
As per Eqs. (\ref{eq:Az}) 
and (\ref{eq:chi}), $\tilde{A}^z_x$ is odd with respect to $y$, 
thus the cross-term $\left\{ \hat{p}_x , \tilde{A}^z_x \sigma_z \right\}/2m_x$ 
averages out upon integration over $y$. As a result, 
upon switching to the dimensionless coordinate $\tilde{x} = \frac{x}{R}$, 
the Hamiltonian $\tilde{\mathcal{H}}_x$ reads 
 \be
 \label{eq:Hx}
\tilde{\mathcal{H}}_x
  = 
  \frac{\hbar^2}{2 m_x R^2}
  \left[ 
  - \frac{d^2}{d \tilde{x}^2} 
 + 
 \frac{[ \pi/2 ]^2}{(1 + \tilde{x}^2)^3} 
 \right] 
   - 
 \frac{\Delta}{2} 
  \frac{ \sin^2 
 \left[ \frac{\pi \tilde{x}}{\sqrt{1 + \tilde{x}^2}} \right] }{\tilde{x}^2}  .
 \ee
 Taken alone, the last term above is obviously beyond the low-energy approximation. 
 However, with decreasing $R$, the repulsion  grows relative to attraction, 
 and overcomes it at an $\bar{R} \sim \sqrt{\frac{m_y}{m_x}} \sqrt{\xi a} \ll \xi$. 
 Thus, Hamiltonian (\ref{eq:Hx}) is valid only in a narrow range 
 $R - \bar{R} \ll \bar{R}$, where the bound state becomes shallow 
 to disappear at $R = \bar{R}$. Notice that $\bar{R} \gg a$ 
 as long as the ratio $\frac{m_y}{m_x}$ is not too small 
 ($\frac{m_y}{m_x} \gg \frac{\Delta}{\epsilon_F}$). 
 
Note that neither the bound state becoming shallow in a narrow range near 
$\bar{R}$ nor its disappearance at $\bar{R}$ rely on the mass anisotropy: 
the same behavior obtains for a perfectly isotropic mass, where the 
Hamiltonian can be diagonalized by solving a single equation for the 
radial wave function. 

Described in this Section, the bound-state behavior at large and 
small Skyrmion radii has several implications. These are convenient 
to illustrate at half-filling, where the total number of states (itinerant 
plus bound) generated by the valence band in the presence of the 
Skyrmion equals the number of electrons -- that is, the number of 
unit cells in the sample. By particle-hole symmetry, every bound state 
$\epsilon_\alpha^c (R)$, split off the conduction band, has a 
valence-band counterpart $\epsilon_\alpha^v (R) = - \epsilon_\alpha^c (R)$, 
as shown in Fig. \ref{Fig:Scales}. Therefore, for any $R$, the 
number of negative-energy states equals the number of electrons.  
Hence, at zero temperature, all the positive-energy states are empty 
while all the negative-energy states are filled. 
Last but not the least, notice that the negative-energy bound states rise 
above the top of the valence band, thus \textit{increasing} the energy 
cost of the Skyrmion. 
Qualitatively, the energy $\epsilon_0^v (R)$ of the highest 
filled state behaves as sketched in Fig. \ref{Fig:Scales}. 
 
\section{Discussion and conclusions} 
\label{section:discussion}

The Skyrmion-electron bound states of Section \ref{section:BPSkyrmion} 
owe their existence to the `texture-induced Zeeman' part of the  
spin-orbit coupling, the last term of Hamiltonian~(\ref{eq:low-E-Hamiltonian-point-X}). 
Unlike the `gauge' term, arising from the Peierls substitution in the kinetic energy 
\cite{Volovik-1987,Tatara-2008}, this term has no equivalent in a ferromagnet. 
At the same time, it hinges on the lower symmetry of $\Sigma$ 
points in the Brillouin zone: at the corner points $X$ or at the center 
point $\Gamma$ in Fig. \ref{Fig:BZ}, such a term is not allowed. 

Notice that, in the effective-mass Hamiltonian of Pekar 
and Rashba, the spin-orbit coupling is linear in momentum 
(see Eqs. (4) and (5) of Ref.  \cite{Pekar-1964}), as is the `gauge' 
spin-orbit term in Hamiltonian (\ref{eq:low-E-Hamiltonian-point-X}). 
By contrast, the `texture-induced Zeeman' spin-orbit coupling does 
not involve electron momentum, and hence is qualitatively different.  

It is instructive to compare the latter term with other known spin-orbit couplings. 
To gain perspective, recall that the textbook Pauli spin-orbit coupling appears 
in the Schr\"odinger Hamiltonian only as a relic of relativity, in second order of 
the expansion in the inverse speed of light  $1/c$ \cite{LL-IV}. 
Remarkably, in antiferromagnets subject to magnetic field, spin-orbit coupling 
may appear in \textit{first} order in $1/c$, via a substantial momentum 
dependence of the $g$-tensor \cite{RR-2021,RR-2008,RR-2009}. 
In contrast to all of the above, the texture-induced spin-orbit coupling 
of Eq. (\ref{eq:low-E-Hamiltonian-point-X}) does not involve $1/c$, 
or the fine-structure constant 
$\alpha = \frac{e^2}{\hbar c} \approx 1 / 137$, at all. 
Instead, the energy scale of the `texture-induced Zeeman' 
spin-orbit term ${v ({\bf A}^\|_y \cdot {\bm \sigma}) \sim \hbar v / L}$ 
is defined by the relevant length scale $L$ of the texture 
and by the Fermi velocity ${v \sim W a / \hbar}$, with $W$ 
the electron bandwidth and $a$ the lattice spacing. 
For the Belavin-Polyakov Skyrmion we analyzed in Section 
\ref{section:BPSkyrmion}, $L$ is given by the Skyrmion radius $R$. 
Compared with $W$, the `texture-induced Zeeman' term is small 
only in the measure of  $L$ being large against $a$: 
${v ({\bf A}^\|_y \cdot {\bm \sigma}) \sim W \frac{a}{L}}$ 
(recall that continuum description is limited to $L \gg a$).  
Relative to the gap $\Delta$, the term is small in the measure of 
$L$ being large against the N\'eel coherence length $\xi = \frac{\hbar v}{\Delta}$, 
that is ${v ({\bf A}^\|_y \cdot {\bm \sigma}) \sim \Delta \frac{\xi}{L}}$.

The Skyrmion-electron bound states are nondegenerate. 
Being limited by the gap, their energy scale is given by $\Delta$. 
In this regard, the texture-induced spin-orbit coupling in 
Eq. (\ref{eq:low-E-Hamiltonian-point-X}) is a real-space analogue of large 
band-splitting effects, discussed for certain types of antiferromagnets \cite{Pekar-1964,Hayami-2019,Yuan-2020,Reichlova-2020,Smejkal-2022}.

We have shown that, in a N\'eel antiferromagnet with certain locations  
of the electron band extrema, a Skyrmion produces non-degenerate 
electron bound states. In each $\Sigma$-valley, low-lying bound states 
are spin-polarized as shown in Fig. \ref{Fig:BZ}. By virtue of charge 
neutrality, at half-filling the bound states do not produce a charge density 
modulation. However, doping the half-filled antiferromagnetic insulator 
by an extra carrier turns the Skyrmion into a charged particle. 
This effect does not rely on the BP profile we used as an illustration, and 
appears for \textit{any} credible shape such as that of a domain-wall Skyrmion. 

Finally, a brief comment on how disorder may limit the validity of our results. 
A crude bound becomes evident upon comparing the energy $\hbar v / R$ 
of the low-lying bound states with the disorder-induced scattering rate $1/\tau$ 
of an electron: 
for a BP Skyrmion of radius $R \gg \xi$, the leading term 
in Eq. (\ref{eq:low-E-at-large-R}) is valid as long as the 
electron mean free path $l = v \tau$ is large compared with $R$. 

Pekar and Rashba showed \cite{Pekar-1964} how a non-uniformity 
of magnetization couples electron spin to its orbital motion. 
However, the analysis of the preceding Sections involved 
no magnetization at all -- not even locally. 
The only quantity present was the staggered (${\bf Q} = (\frac{\pi}{a} , \frac{\pi}{a})$) 
magnetization and its inhomogeneity. Thus, Pekar's and Rashba's insight holds 
beyond their original statement: in a \textit{general} magnetically ordered system, 
inhomogeneity begets spin-orbit coupling. 

Skyrmion-electron bound states are a new arrival in the family 
of electron states, localized on topological defects such as dislocations 
\cite{Landauer-1954}, vortices in superconductors \cite{Caroli-1964} 
or solitons in organic materials \cite{Brakir-1984,Heeger-1988}. 
Becoming charged in the presence of dopant carriers,  
the Skyrmion can be manipulated by electric field, which may open 
new possibilities for its use in devices. We hope that our results stimulate 
further work both on fundamental and applied aspects of this phenomenon. 

\acknowledgments{
We thank P. Pujol for the many discussions and helpful suggestions. 
We are grateful to Ya. B. Bazaliy, M. V. Kartsovnik and A. Monin for 
illuminating comments. Finally, we thank G. Baskaran for pointing us, 
after this work appeared on the arXiv, to early studies 
\cite{John-Golubentsev-1993,John-Golubentsev-1995} 
that found Skyrmion-electron bound states in an exotic 
spin liquid (see Appendix \ref{section:comparison}).
}

\appendix

\section{When can one treat the Belavin-Polyakov Skyrmion as static?} 
\label{section:static}

For $R \gtrsim \xi$, treating the BP Skyrmion profile as static input 
to the electron problem requires the bound-state level spacing 
$\hbar \omega_e \sim \Delta \left( \frac{\xi}{R}\right)^{\frac{3}{2}}$ 
in Eq. (\ref{eq:low-E-at-large-R}) to be large against the energies 
of high-frequency localized eigenmodes of the Skyrmion. These 
are bound from above by the spin wave gap $\hbar \Omega_0$ 
of the bulk magnon spectrum \cite{Kravchuk-2019}. In the relevant 
limit of  small single-ion anisotropy $K \ll J$, the behavior 
$\hbar \Omega_0 \sim \sqrt{KJ}$ \cite{Kittel-1951,Rezende} 
translates the condition $\omega_e \gg \Omega_0$ into 
$\Delta \left( \frac{\xi}{R}\right)^{3/2} \gg \sqrt{KJ}$. Which means 
that a sufficiently large BP Skyrmion can no longer be treated as static. 
For a weak enough anisotropy, this limits our treatment to radia 
\be
\label{eq:R-static}
R \ll \xi \left(\frac{\Delta}{J}\right)^{2/3} \left(\frac{J}{K}\right)^{1/3} .
\ee
Inequality (\ref{eq:R-static}) is meaningful only 
if its r.h.s. is large compared with $\xi$, that is if 
\be
\label{eq:K-static}
\frac{K}{J} \ll \left(\frac{\Delta}{J}\right)^2 .
\ee

Now, the conditions (\ref{eq:R-static}), (\ref{eq:K-static}) define 
the possibility of treating the Skyrmion as static when dealing 
with $y$, the `fast' coordinate of the electron. As per 
Eq. (\ref{eq:low-E-at-large-R-full}), for the 
`slow' coordinate, $x$, the relevant frequency is 
$\hbar \omega_e^x \sim \Delta \sqrt{\frac{m_y}{m_x}} 
\sqrt{\frac{a}{\xi}} \left( \frac{\xi}{R}\right)^{\frac{3}{2}}  
\ll \hbar \omega_e$. As a result, the condition for treating the 
Skyrmion as static when dealing with $x$ is more stringent 
than (\ref{eq:R-static}):
\be
\label{eq:R-static-x}
R \ll \xi \left( \frac{\Delta}{\epsilon_F} \right)^{1/3} 
\left( \frac{m_y}{m_x} \right)^{1/3} 
\left(\frac{\Delta}{J}\right)^{2/3} \left(\frac{J}{K}\right)^{1/3} .
\ee
Inequality (\ref{eq:R-static-x}) makes sense only 
if its r.h.s. is large compared with $\xi$, that is if 
\be
\label{eq:K-static-x}
\frac{K}{J} \ll \left(\frac{\Delta}{J}\right)^2 \frac{m_y}{m_x} \frac{\Delta}{\epsilon_F}.
\ee
Put otherwise, the levels emerging from quantizing the electron motion 
along the `slow' coordinate $x$ define the `fine' structure of the bound-state 
spectrum as opposed to the `gross' structure arising from quantization along 
the `fast' coordinate $y$. 
This `fine' structure can be treated in the static-Skyrmion approximation 
under conditions (\ref{eq:R-static-x}), (\ref{eq:K-static-x}) that are, 
naturally, much more stringent than similar inequalities 
(\ref{eq:R-static}), (\ref{eq:K-static}) for the `gross' structure 
of Eq. (\ref{eq:low-E-at-large-R}). 
At the same time, no matter how small the $\omega_e/\Omega_0$ 
ratio, a sufficiently large number of Skyrmion eigen-excitations 
would eventually influence the excited electron bound states 
(at an appropriately high order of perturbation theory). 

\section{Earlier work 
\cite{John-Golubentsev-1993,John-Golubentsev-1995,Cheng-2012,Haas-1996,Morinari-2012,Shraiman-1988}} 
\label{section:comparison} 

In this section of the Appendix, we discuss some of the relevant early work.

S. John and A. Golubentsev \cite{John-Golubentsev-1993,John-Golubentsev-1995} 
studied the Hubbard model on a square lattice in a topological spin liquid state, 
defined by two key properties: (i) checkerboard N\'eel order, and 
(ii) anti-periodicity of the electron wave function along closed path around 
any elementary plaquette of the lattice. As a consequence, the unit cell 
in such a state quadruples relative to the underlying square lattice, and 
the electron wave function is thus a 4-spinor. The model band extrema 
fall at the very same points $\Sigma = (\pm \frac{\pi}{a} , \pm \frac{\pi}{a})$ 
that we focussed on, and all the four points $\Sigma$ are equivalent in the 
`ordered' Brillouin zone. The resulting `Dirac' electron spectrum near $\Sigma$ 
is thus isotropic by symmetry. 

As opposed to the above, in the N\'eel state of our interest the unit cell doubles 
rather than quadruples and, therefore, the electron wave function is a bispinor 
rather than a 4-spinor. 
Contrary to being all equivalent in the spin liquid state, in the N\'eel phase 
the four points $\Sigma$ split into two inequivalent pairs $\Sigma_{1,2}$ 
and  $\Sigma_{3,4}$, shown in Fig. \ref{Fig:BZ}. In contrast to the spin-liquid state, 
the symmetry of the N\'eel state does not require the electron spectrum 
near the $\Sigma$ points to be isotropic. Quite to the contrary, it tends 
to be strongly anisotropic, consistently with experimental findings 
\cite{Armitage,Matsui,Song} in a number of cuprates. 

Lastly, the Skyrmion-electron bound states we found are non-degenerate. 
In each of the four $\Sigma$-valleys, low-lying bound states are spin-polarized 
as shown in Fig. \ref{Fig:BZ}, whereas the bound states of Refs. 
\cite{John-Golubentsev-1993,John-Golubentsev-1995} are doubly 
degenerate as a consequence of the elevated symmetry of the 
spin-liquid state. 

To summarize, the similarities between the Skyrmion-electron bound states 
of our work and those found by John and Golubentsev arise from N\'eel order 
being an ingredient of the spin liquid studied in Refs. 
\cite{John-Golubentsev-1993,John-Golubentsev-1995}. 
Such similarities include the shallow character of bound states at 
large BP Skyrmion radia, although John and Golubentsev have 
not explored the low-energy limit, instead opting for solving their 
full 8$\times$8 Hamiltonian numerically. 

The differences between our results and those of John and Golubentsev 
stem from the features of their spin liquid that are not inherent to a generic 
N\'eel state that we studied, such as (i) an elevated symmetry of the spin liquid, 
and (ii) the wave function antiperiodicity under translation around 
any elementary plaquette of the square lattice. Such differences 
lead to double degeneracy of the bound states found by John and 
Golubentsev, while ours are non-degenerate. 

Now we turn to the work \cite{Cheng-2012} by R. Cheng and Q. Niu, 
who derived coupled quasiclassical equations of motion for the electron momentum, 
coordinate and spin in an antiferromagnetic texture. By construction, such a 
description holds only for large quantum numbers and, in fact, the equations 
of Cheng and Niu become singular at the MBZ boundary ($\xi \rightarrow 0$), 
see the r.h.s. of Eq. (8c) of Ref. \cite{Cheng-2012}. 
By contrast, our effective electron Hamiltonian 
(\ref{eq:low-E-Hamiltonian-general}-\ref{eq:low-E-Hamiltonian-point-X}) 
covers both the quasiclassical regime and the extreme quantum limit 
as long as the electron energies are close to the gap edge 
($\frac{|E - \Delta|}{\Delta} \ll 1$).

Finally, we would like 
to mention the work \cite{Haas-1996,Morinari-2012} 
on different versions of the $t$-$J$ model of high-temperature superconductivity. 
These studies were performed on small clusters in the atomic limit, with exchange 
integrals substantial or even large compared with the hopping matrix elements.  
The authors of both Refs. \cite{Haas-1996,Morinari-2012} reach a conclusion 
that is, in a way, reciprocal to ours (``a Skyrmion produces electron bound states''): 
namely, that introducing a dopant carrier into a single CuO plane renders 
a Skyrmion \cite{Haas-1996} or a half-Skyrmion \cite{Morinari-2012} 
configuration energetically favorable \cite{Shraiman-1988}. 
When applied to a single dopant carrier bound to a Skyrmion, this conclusion 
stems from neglecting the valence band giving rise to filled bound states, 
whose energy \textit{increases} rather than decreases (see Fig. \ref{Fig:Scales}), 
and thus overwhelms the energy gain found in Refs. \cite{Haas-1996,Morinari-2012}. 
A more technical difference with respect to our work is that the studies \cite{Haas-1996,Morinari-2012} were performed in the atomic limit, 
and thus produced Skyrmion sizes of the order of the lattice spacing, 
where the continuum description is not applicable.


\begin{thebibliography}{9}
\bibitem{footnote:re-Hamiltonian} 
The terms omitted when passing from Eq. (\ref{eq:low-E-Hamiltonian-general}) 
to (\ref{eq:low-E-Hamiltonian-point-X}) are higher order in momentum. 
They are negligible as long as the low-energy approximation of 
Eqs. (\ref{eq:low-E-Hamiltonian-general}-\ref{eq:low-E-Hamiltonian-point-X}) 
is valid.
\bibitem{footnote:x-reflection} 
Symmetry under reflection $x \rightarrow -x$ is imposed by the small group 
of point $\Sigma$, and thus forbids the appearance of gauge-invariant terms 
such as $( A^\|_x \cdot A^\|_y )$ and $\left[ A^\|_x \times A^\|_y \right]$.
\bibitem{footnote:m_x} 
Notice that, with only the nearest-neighbor hopping, 
$m_x$ at $\Sigma$ is infinite.
\bibitem{footnote:BP}
The strikingly simple r.h.s. of Eq. (\ref{eq:vF}) 
is a remarkable property of the BP Skyrmion.
\bibitem{footnote:small} 
This conclusion can be reached by direct comparison, 
with the single exception of the cross-term 
$V = \left\{ p_y , A_y^z \sigma_z \right\}/2 m_y^*$. 
For the latter, the presence of $\sigma_z$ implies that  this 
term has matrix elements only across the gap, and thus its 
contribution is suppressed by the perturbation-theory 
factor of  the order of $V  / \Delta \ll 1$.
\bibitem{footnote:chirality} 
For chirality $\gamma \neq 0$, spin polarization of the bound state 
is tilted by $\gamma$ relative to the one shown in Fig. \ref{Fig:BZ}.
\end{thebibliography}
\end{document}